\documentclass[aps,prl,reprint,superscriptaddress]{revtex4-2}
\usepackage{graphicx}
\begin{document}

%\preprint{}

\title{\textbf{Orbital-interaction-aware Deep Learning Model for Efficient Surface Chemistry Simulations}}


\author{Zhihao Zhang}
\affiliation{State Key Laboratory of Green Chemical Engineering and Industrial Catalysis, center for Computational Chemistry and Research Institute of Industrial Catalysis, East China University of Science and Technology, Shanghai 200237, P. R. China}
\author{Xiao-Ming Cao}
\email[Contact author:]{xmcao@sjtu.edu.cn}
%\thanks{}
\affiliation{State Key Laboratory of Green Chemical Engineering and Industrial Catalysis, center for Computational Chemistry and Research Institute of Industrial Catalysis, East China University of Science and Technology, Shanghai 200237, P. R. China}
\affiliation{State Key Laboratory of Synergistic Chem-Bio Synthesis, School of Chemistry and Chemical Engineering, Shanghai Jiao Tong University, Shanghai 200240, P. R. China}


\date{\today}

\begin{abstract}
Deep learning has advanced efficient chemical process simulations on the surfaces, accelerating high-throughput materials screening and rational design in heterogeneous catalysis, energy storage and conversion, and gas separation. However, the accuracy of the deep learning model generally depends on the quality of the training data. Unfortunately, precise experimental data in surface chemistry, such as adsorption energies, are scarce, while accurate quantum chemistry simulations remain computationally prohibitive for large-scale studies. Herein, we present a deep learning model of DOS Transformer for Adsorption (DOTA) for efficient surface chemistry simulations with chemical accuracy. It enables the alignment of experimental data and multi-fidelity quantum chemistry calculation data by capturing latent orbital interaction patterns based on the map between local density of states (LDOS) and adsorption energy. This minimizes the reliance on scarce high-precision training data in surface chemistry to accomplish efficient prediction of adsorption energies rivaling the high-precision experimental data, resolving the long-standing challenge of "CO puzzle". It provides a robust framework for efficient materials screening, effectively bridging the gap between computational and experimental data.
\end{abstract}

%\keywords{}

\maketitle

%\section{Introduction}
%\textit{Introduction---}
Molecular adsorption and desorption are indispensable for the surface chemical processes in heterogeneous catalysis, energy storage and conversion, and gas capture. This makes the achievement of precise adsorption energy ($E_{\rm{ad}}$) essential for the accurate understanding of surface chemical processes and the rational design of improved materials within these applications \cite{1_2,1_4,1_7,4_5}. Due to the substantial inherent uncertainties arising from side reactions, surface defects, impurities, etc., precise experimental $E_{\rm{ad}}$ are mainly obtained through single-crystal adsorption microcalorimetry with an elaborate experimental design so far, resulting in scarce benchmark experimental data \cite{2_1,2_2,2_3,2_4,2_7,2_9}. The quantum chemistry simulation is an alternative to achieving accurate $E_{\rm{ad}}$. However, it faces the cost-accuracy trade-off. High-ladder CCSD(T) and random phase approximation (RPA) \cite{13_1,13_3,13_4}, are computationally expensive for massive surface simulations, while density functional theory (DFT) calculations with lower computational costs, especially for generalized gradient approximation (GGA) exchange-correlation functionals \cite{10_1}, exhibit limitations in terms of accuracy, exemplified by the classic "CO puzzle" problem \cite{11_1,11_2,11_3,11_4,11_5} for the widely used PBE functional \cite{PBE}. Moreover, the meta-GGA functionals of SCAN \cite{SCAN} and r$\rm ^2$SCAN \cite{r2SCAN}, which have been regarded as promising functionals to establish the reliable $E_{\rm{ad}}$ database, even overbind slightly compared to PBE and mistakenly predict the preferential site for CO adsorption as well \cite{12_1}. Notably, even with DFT calculations, the computational cost remains unaffordable for high-throughput $E_{\rm{ad}}$ calculations. This limits the application in screening candidate materials. Deep learning models have recently emerged as a promising approach to accelerate high-throughput surface chemistry simulations, along with $E_{\rm{ad}}$ \cite{4_1,4_2,4_4,4_8,4_9}. Specifically, machine learning interatomic potentials (MLIP), which obtain optimized adsorbate structures and adsorption energies using structure coordinates as inputs, have been established as a critical technique in recent years \cite{6_1,6_2,6_4,6_5}. \par
 
 However, accurately predicting $E_{\rm{ad}}$ using a deep learning model suffers from the scarcity of high-precision training data. In general, the accuracy of the $E_{\rm{ad}}$ prediction is subject to the accuracy of the training set. The constraint of computational cost renders the adsorption databases \cite{8_1,8_2,8_3} predominantly composed of the results based on GGA functionals so far. This hinders the prediction of $E_{\rm{ad}}$ rivaling the high-precision experimental results. Recently, the $\rm \Delta$-machine learning method presents an alternative approach for predicting $E_{\rm{ad}}$ with chemical accuracy for specific systems. Nevertheless, approximately one hundred RPA results are still required to train a model to predict the CO adsorption energy ($E_{\rm{ad}}$(CO)) on a specific metal surface using the $\rm \Delta$-machine learning method \cite{14_1,14_2,14_3,14_4}. Consequently, tremendous computational expenditures limit the extension of the $\rm \Delta$-machine learning method to vast chemical compound spaces. It is imperative to develop a new model and protocol to predict $E_{\rm{ad}}$ with chemical accuracy, leveraging the multi-fidelity DFT data and scarce high-precision experimental or theoretical simulation $E_{\rm{ad}}$ data. \par

\begin{figure} 
	\includegraphics{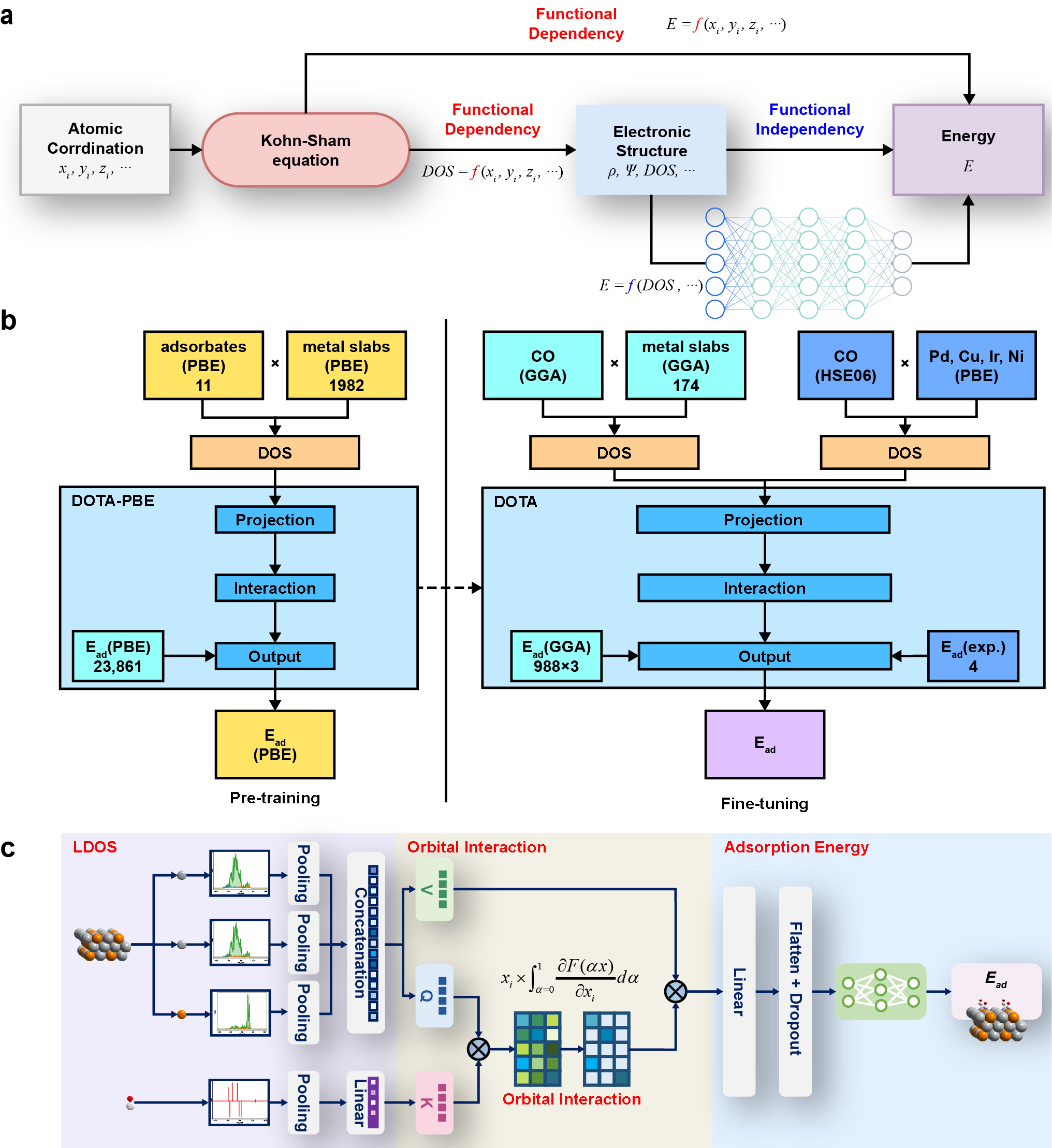} 
	\caption{\label{Fig. 1}a) Procedure from coordination to energy. b) Pretraining and fine-tuning workflow of DOTA exemplified by CO adsorption. c) Illustration of DOTA, with PDOS partition in the lavender area, orbital interaction in the yellow area, and output $E_{\rm{ad}}$ in the blue area.}
\end{figure}

To this end, aligning multi-fidelity data is a prerequisite for a deep learning model allowing for chemically precise $E_{\rm{ad}}$ prediction. The predicted $E_{\rm{ad}}$ by MLIP, which corresponds to the mapping from input atomic coordinates to output energies, must be theory-dependent or functional-dependent in the case of using DFT-labeled data. This arises from the intrinsic functional independence of atomic coordinates, while the functional dependency of energies (\textbf{Fig. ~\ref{Fig. 1}a}). This indicates that a functional-independent deep learning model can be achieved by employing functional-dependent physical quantities as both model input and output. Notably, both $E_{\rm{ad}}$ and electronic structure characteristics exhibit strong functional dependency. Hence, it is possible to unify different-fidelity data, including experimental and different-level quantum chemistry data in surface chemistry, upon the map between electronic structures and $E_{\rm{ad}}$. The density of states (DOS) serves as an information-dense descriptor that compresses a substantial amount of spatially distributed electronic structure information into a computationally affordable one-dimensional energy space, making it a promising candidate as a model input to overcome the limitation of functional dependency. Moreover, DOS can help to understand the chemisorption strength in heterogeneous catalysis \cite{16_1, 19_1, 18_1,20_1}. \par

In this spirit, we present DOS Transformer for Adsorption (DOTA) model for efficient $E_{\rm{ad}}$ predictions to rival with high-precision experimental results. The DOTA architecture synergistically integrates the interpretable multi-head self-attention mechanism of the Transformers framework with local DOS (LDOS) feature engineering to comprehensively understand the contribution of different energy levels to $E_{\rm{ad}}$. The underlying orbital interaction patterns are captured by the pretraining upon the low-cost PBE-level LDOS profiles and $E_{\rm{ad}}$ without prior physical constraints. Based on the generality of these orbital interaction patterns across different quantum chemistry methods and experimental data, the $E_{\rm{ad}}$ with chemical accuracy of the studied adsorbate across various metallic and intermetallic surfaces could be predicted by fine-tuning the PBE-level model with the incorporation of the LDOS profile of the studied gaseous adsorbate at the HSE06 \cite{HSE} level and single-digit high-precision $E_{\rm{ad}}$ for each adsorbate. 

%\section{Results}
%\textit{Results---}
%\subsection{Workflow of DOTA}
%Workflow of DOTA:\quad
To enable efficient catalyst screening without requiring preliminary time-consuming geometry optimization during the inference, DOTA strategically processes gaseous adsorbate LDOS and bare surface LDOS instead of adsorbed complex configurations. The model architecture (\textbf{Fig. ~\ref{Fig. 1}c}) implements quantum mechanically informed feature engineering, allocating 32 angular momentum-projected DOS (PDOS) embedding channels per surface atom and 8 PDOS embedding channels per adsorbate atom involved in the bonding formation.  This channel design ensures compatibility with f-electron systems and spin-polarized calculations. Initial feature processing involved an average pooling layer to normalize energy interval disparities across different DOS calculations. The multi-head attention mechanism enables efficient modeling of long-range orbital interactions across the entire energy spectrum, capturing the interactions between orbitals with different angular momentum quantum numbers. Final energy predictions are generated by merging surface atom features through flattening and sequential fully connected layer processing. (see SM for more details \cite{SM}) \par

\begin{figure*}
	\includegraphics{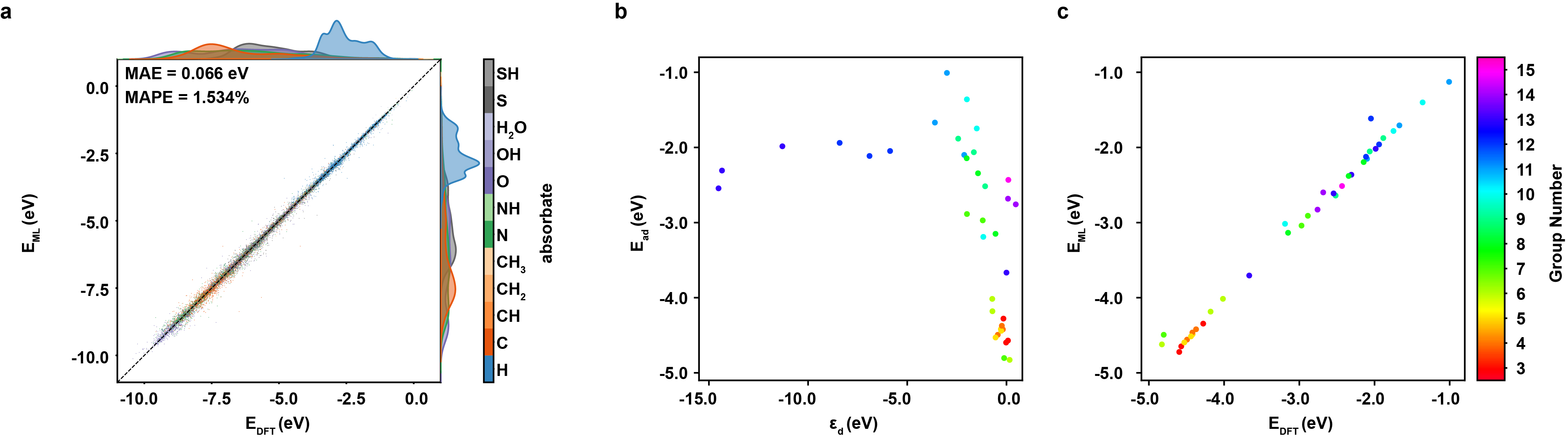} 
	\caption{\label{Fig. 2}a) Parity and kernel density plots of $E_{\rm{ad}}$ between PBE calculations and DOTA-PBE from five-fold cross-validation. b) Plot of PBE-calculated $E_{\rm{ad}(OH)}$ vs. d-band center. c) Plot of DOTA predicted $E_{\rm{ad,PBE}}(\rm{OH})$ vs. PBE-calculated $E_{\rm{ad}}(\rm{OH})$. }
\end{figure*}

The DOTA framework employs a two-stage training protocol comprising pretraining and fine-tuning steps to predict chemically precise $E_{\rm{ad}}$ (\textbf{Fig. ~\ref{Fig. 1}b}). During the pretraining step, all the LDOS inputs and corresponding $E_{\rm{ad}}$ outputs were derived from PBE results, thereby constructing a DOTA-PBE model for $E_{\rm{ad,PBE}}$ prediction. The pretraining dataset encompasses monoatomic adsorbates (H, C, N, O, and S) and their hydrogenated counterparts (CH, $\rm CH_2$, $\rm CH_3$, NH, OH, $\rm H_2O$, and SH) across 1,982 distinct metallic and intermetallic (111) surfaces (\textbf{Table S2}). The selected adsorbates cover molecules with varying degrees of bonding unsaturation, enabling $E_{\rm{ad}}$ to span from near 0~eV to over -10~eV. The common latent orbital interaction patterns between adsorbates and metallic surfaces are expected to be learned through the pretraining based on these PBE data. The predictive model for the chemically precise $E_{\rm{ad}}$ was subsequently achieved by fine-tuning the DOTA-PBE model by multi-head strategy. One fine-tuning dataset is composed of the GGA-level LDOS-$E_{\rm{ad}}$ data pairs. It is used to learn the specific orbital interaction patterns for the studied adsorbate. The other fine-tuning dataset only includes single-digit hybrid data pairs. Each hybrid data pair consists of the input HSE06-level DOS profile of the gaseous adsorbate molecule (LDOS(g, HSE06)), the input PBE-level LDOS profile of the bare surface (LDOS(surf, PBE)), and the corresponding high-precision $E_{\rm{ad, exp.}}$ as the output. \par

%\subsection{Performance of DOTA-PBE}
%Performance of DOTA-PBE:\quad
\textbf{Fig. ~\ref{Fig. 2}a} shows that the DOTA-PBE model achieves exceptional accuracy for $E_{\rm{ad,PBE}}$ prediction with a mean absolute error (MAE) of 0.066~eV and a mean absolute percentage error (MAPE) of 1.53~$\rm \%$ across all the adsorption systems. Notably, the reliance on electronic structure features rather than geometric parameters enables effective transferability between adsorbates with analogous bonding characteristics. \par

\begin{figure}
	\includegraphics{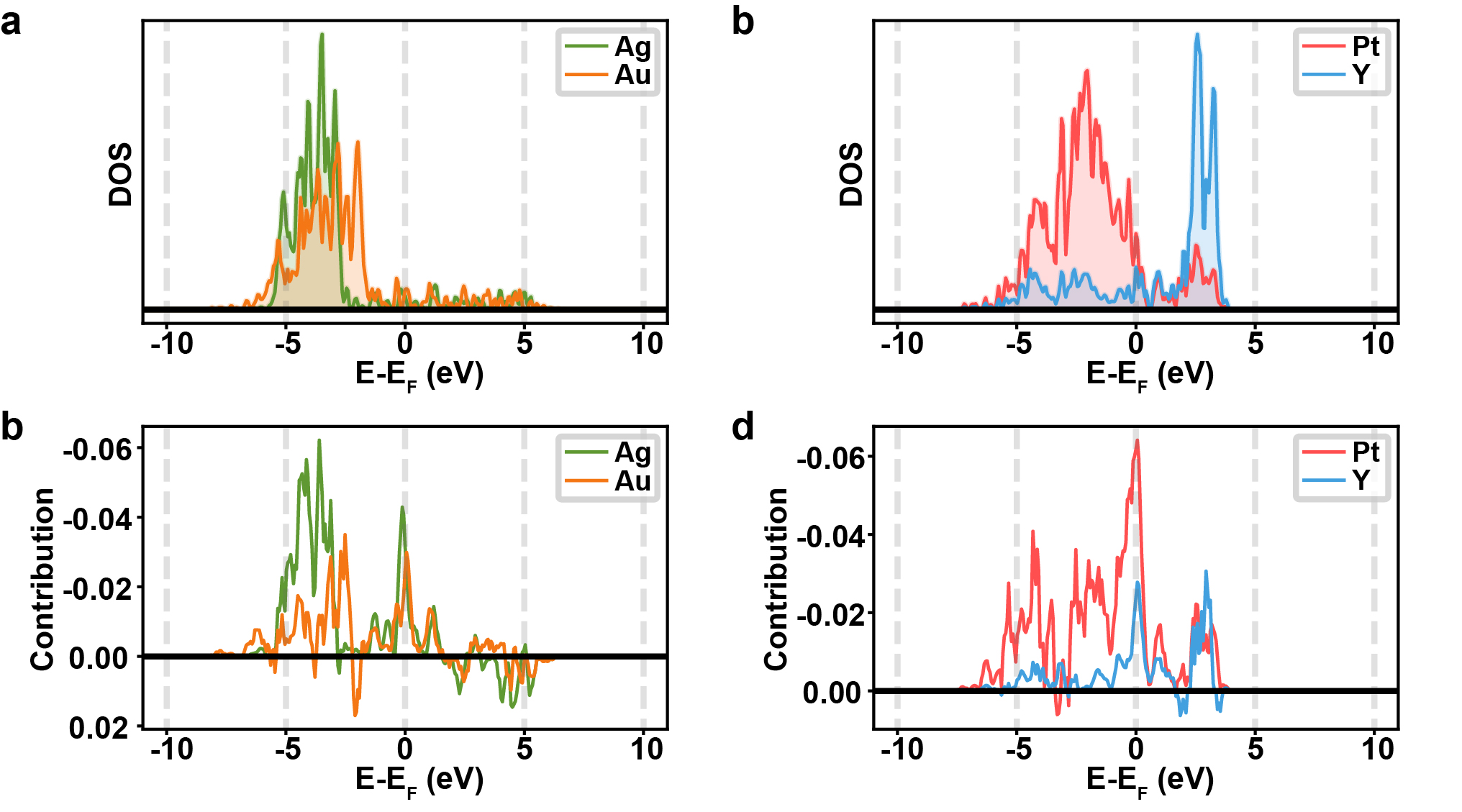} 
	\caption{\label{Fig. 3}a) LDOSs of surface Ag in Ag(111) and surface Au in Au(111), b) the contribution of each energy level of surface Ag in Ag(111) and surface Au in Au(111) to the OH adsorption atop Ag(111) and Au(111).  c) LDOSs of surface Pt and Y atoms in $\rm Pt_3Y$(111).  d) the contribution of each energy level of surface Pt and Y to H adsorption atop $\rm Pt_3Y$(111).}
\end{figure}

We further employed the interpretability of our framework to revisit the d-band center \cite{16_1} and Fermi softness models \cite{19_1}. The OH adsorption atop unseen (111) surfaces in DOTA-PBE training (\textbf{Fig. ~\ref{Fig. 2}b}) indicates that the d-band center theory \cite{16_1} fails at the sites composed of the group 13-14 elements that rely on the s and p orbitals for bonding, while the accuracy of the DOTA-PBE model is element-independent across the metallic and intermetallic surfaces (\textbf{Fig. ~\ref{Fig. 2}c}). The d-band center theory also fails in Ag(111) and Au(111) surfaces. The lower d-band center of Ag(111) ($\rm \varepsilon_d$ = -3.61~eV), compared with the Au(111) surface ($\rm \varepsilon_d$ = -3.00~eV), instead leads to the stronger $E_{\rm{ad,PBE}}$ of OH atop Ag(111) (-1.67~eV) than Au(111) (-1.01~eV). Moreover, the d-band of the Ag(111) surface is mainly concentrated in the energy window far below the Fermi level, implying weak adsorption if following Fermi softness theory. The decomposition of the contribution of each energy level (\textbf{Fig. ~\ref{Fig. 3}b}) by the integrated gradients algorithm \cite{22_1} suggests that those energy levels far below the Fermi level made the main contribution for OH adsorption atop Ag(111) rather than those energy levels around the Fermi level. The electron transfer from the metallic surface to the singly occupied molecular orbital of OH \cite{29_1} or Pauli repulsion \cite{29_2}, rather than the orbital overlapping between OH and the Ag(111) or Au(111) surface, dominates the chemisorption of OH atop Ag(111) or Au(111), which is difficult to be described by the d-band center or the Fermi softness theory. In contrast, our DOTA-PBE model, which omits any predefined restriction, can predict the $E_{\rm{ad, PBE}}$ of OH atop Ag(111) and Au(111) surfaces, with a prediction of -1.71~eV and -1.13~eV, respectively. \par

The DOTA-PBE model also helps us to understand why that d-band center theory fails to predict the H adsorption site over Pt$\mathrm{_3}$Y(111) \cite{19_1}. PBE calculations show that the $E_{\rm{ad, PBE}}$(H) at the Pt top site is -2.80~eV, while it is -0.81~eV atop the Y site, indicating that H tends to preferentially adsorb at the Pt site. The DOTA-PBE model can also predict this case, with the predicted $E_{\rm{ad, PBE}}$ atop the Pt and Y sites being -2.75~eV and -0.88~eV, respectively. However, the Pt site has a lower d-band center ($\rm \varepsilon_d$ = -0.32~eV) than the Y site ($\rm \varepsilon_d$ = 0.23~eV), indicating stronger adsorption at the Y site if following the d-band center theory. The contribution of each energy level to the H adsorption ($\textbf{Fig. ~\ref{Fig. 3}c}$) shows that the conduction bands of Y play an important role in the H chemisorption atop the Y site. Notably, the upshift of the Y conduction band could lead to the upshift of the d-band center synchronously, thereby predicting a stronger $E\rm_{ad, PBE}$(H) by d-band center theory. However, the upshift of the Y conduction band would widen the energy gap between the H 1s orbital and the Y conduction band, constraining their orbital overlap. Consequently, the higher the d-band center, the weaker the H adsorption atop the Y site. It indicates that the d-band center model is not suitable for describing the system, where the conduction band serves a main role in $E\rm_{ad}$. \par

\begin{figure}
	\includegraphics{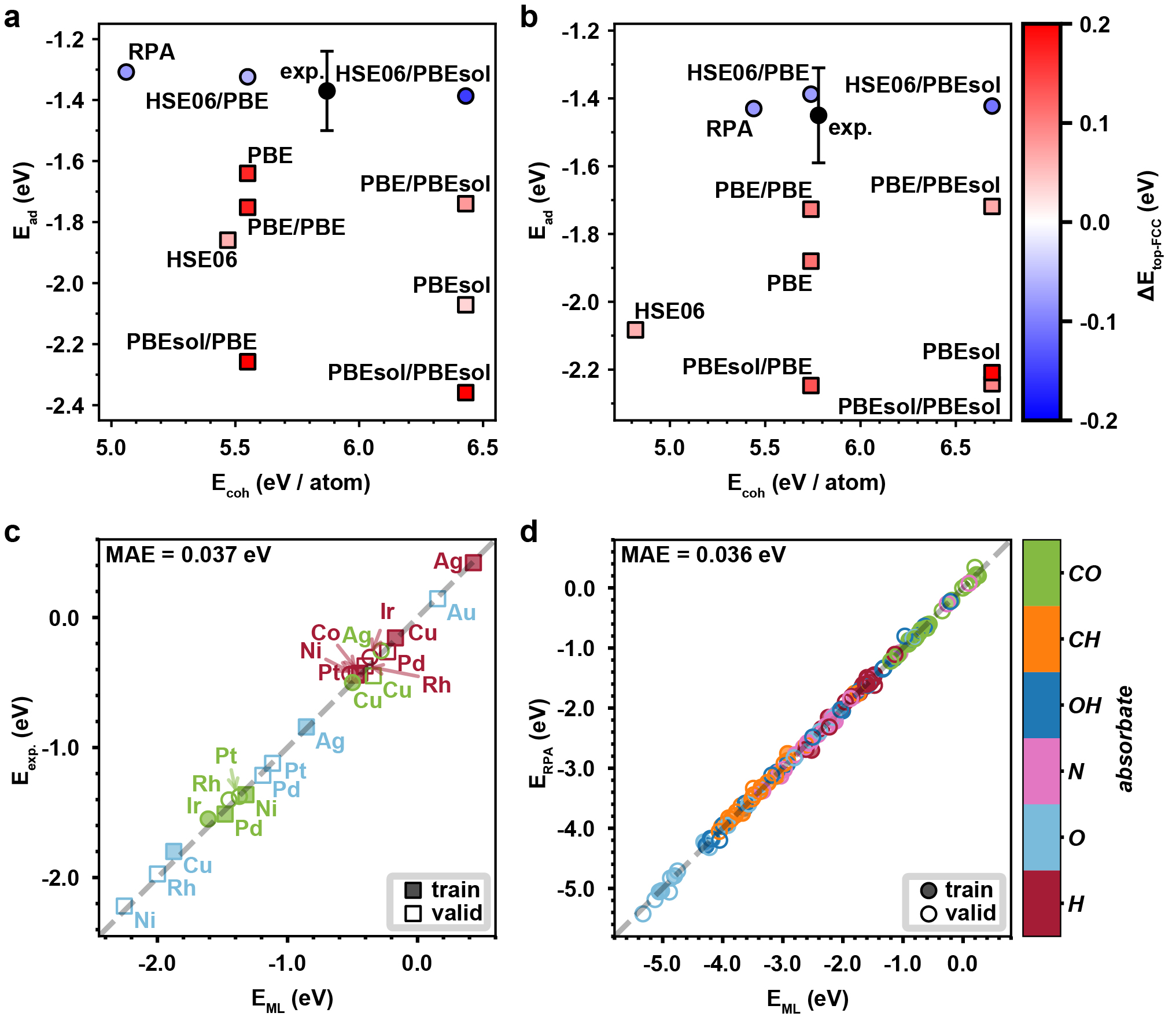} 
	\caption{\label{Fig. 4} Cohesive energies and predicted $E_{\rm{ad}}(\rm{CO})$ using various combinations of functionals a) on Pt(111) and b) on Rh(111). The $E_{\rm{ad}}$ for the preferred site is marked. The circle and square correspond to the preferred adsorption at the top and fcc sites, respectively. The energy difference between the top and fcc sites is indicated by a color gradient. c) Experimental vs. DOTA-predicted $E_{\rm{ad}}(\rm{CO})$ and dissociative $E_{\rm{ad}}$ of $\rm H_2$ and $\rm O_2$. All data points except the Cu(111)-fcc value (emb-MC-PDFT) are experimental values \cite{3_1}. d) DOTA-predicted $E_{\rm{ad}}$ at 1 ML coverage vs. RPA reference data.}
\end{figure}

%\subsection{Predictive chemically accurate adsorption energy}
%Predictive chemically accurate adsorption energy:\quad
Building upon the robust performance and transferability of DOTA-PBE, we fine-tuned the model to predict $E_{\rm{ad}}$ rivaling the high-precision experimental results for specific adsorbates across metallic and intermetallic surfaces. Taking CO adsorption as an example, we first incorporated 988 LDOS-$E_{\rm{ad}}$(CO) pairs (\textbf{Table S2}) across different sites of 174 metallic and intermetallic surfaces, evaluated by PBE, RPBE \cite{RPBE}, and PBEsol \cite{PBEsol} functionals, respectively, as one fine-tuning dataset. Data derived from different functionals expanded the perceptive chemical space, thereby enhancing the ability of the model to learn orbital interaction patterns between CO and metallic surfaces. Importantly, the foundation of PBE-level common orbital interaction understanding, combined with GGA-level insights into the interactions between CO and the surface, enabled exceptional data efficiency to achieve chemical accuracy predictions with the requirement of only four high-precision data points. Each high-precision data point comprised experimentally measured $E_{\rm{ad,exp.}}\rm{(CO)}$ \cite{3_1} as output, paired with corresponding input LDOS(surf, PBE) and LDOS(CO(g), HSE06) characteristics (\textbf{Fig. S1a}). We compared $E_{\rm{ad}}$(CO) over Pt(111) (\textbf{Fig. ~\ref{Fig. 4}a} and \textbf{Table S3}) and Rh(111) (\textbf{Fig. ~\ref{Fig. 4}b} and \textbf{Table S4}) predicted by the DOTA model with different input functional combinations for adsorbate/surface LDOS profiles, the standard PBE, PBEsol, HSE06, RPA, and experimental results. Each prediction using the combination of functionals by DOTA is labeled following the format of the functional used for the adsorbate/the functional for the metal slab. \textbf{Fig. ~\ref{Fig. 4}a} shows that both the standard PBE and PBE/PBE combinations result in CO overbonding and mistakenly predicted fcc sites for CO chemisorption over these two surfaces. More poor results were predicted by the standard PBESol and PBESol/PBESol combinations. Since PBE exhibits good performance at the description of the metallic surface electronic structure, as demonstrated by the closer metal cohesive energy to the experimental value (\textbf{Fig. ~\ref{Fig. 4}a} and \textbf{Fig. ~\ref{Fig. 4}b}), this demonstrated that the significantly underestimated HOMO-LUMO gap of CO by PBE and PBEsol functionals \cite{30_1,30_2} should account for the overbinding of CO over the Pt(111) and Rh(111) surfaces \cite{24_1,24_2}. The standard HSE06 functional is also poor at predicting CO adsorption because it fails to describe the delocalized electronic structures of metallic surfaces \cite{28_1, HSE}. Despite the accurate $E_{\rm{ad}}$(CO), neither the CO HOMO-LUMO gap nor the metal cohesive energies could be accurately calculated by RPA (\textbf{Fig. ~\ref{Fig. 4}a}, and \textbf{Fig. ~\ref{Fig. 4}b}), which is consistent with previously reported results. \cite{11_5} Importantly, HSE06 exhibits the best predictive performance for CO HOMO-LUMO energy levels, while the PBE functional excels at the prediction of metal cohesive energy \cite{11_5}, indicating that the HSE06 and PBE functionals are good at the description of the molecular orbitals and metal surfaces, respectively. Consequently, the input of LDOS(CO(g), HSE06) and LDOS(surf, PBE) (HSE06/PBE) could provide superior results, with the predicted $E_{\rm{ad}}$ differing by less than 0.04~eV of chemical accuracy from the experimental values. More importantly, the model with the HSE06/PBE input combination could correctly predict the top site as the preferential binding site for CO adsorption, aligning with the experimental results \cite{3_1}. This also manifests that the underestimated HOMO-LUMO gap by the PBE functional is mainly responsible for the "CO puzzle" \cite{23_1}. \par

We further examined the transferability of the fine-tuned DOTA framework to dissociative adsorption of $\rm H_2$ and $\rm O_2$. Due to the scarcity of high-fidelity experimental dissociative adsorption energies \cite{Ads_Rh, Ads_Au, Ads_Pt, Ads_Pd, Ads_Cu, Ads_Ir, Ads_Ag, Ads_Ni, Ads_Co}, the model was trained on only 3 data points for H and 2 for O, and validated against 5 independent measurements for each. The $E_{\rm{ad}}$ of the fcc site on Cu(111) was from highly accurate quantum-embedded multi-configuration pair-density functional theory \cite{Ads_CuFCC}. \textbf{Fig. ~\ref{Fig. 4}c} shows that predictions for both training and validation sets cluster closely around the parity line across all three molecules based on minimal supervision. Moreover, the MAE value of 0.037~eV for validation sets is also within chemical accuracy. These preliminary results suggest that DOTA’s orbital-interaction-aware architecture enables robust few-shot learning not only for molecular adsorption but also for dissociative adsorption, highlighting its potential for modeling more surface reactions such as hydrogen evolution or oxygen reduction.\par

The DOTA model could be generalized to cases for the adsorbates with different coverages. We fine-tuned the pretrained DOTA-PBE model for predicting the accurate $E_{\rm{ad}}$ of six adsorbates (H, O, N, OH, CH, CO) across 25 transition metal surfaces at the 1 ML coverage. The training set for fine-tuning comprised 120 PBE, 120 RPBE, and 48 reported RPA calculations (8 RPA $E_{\rm{ad}}$ for each adsorbate), while the validation set contained 30 PBE, 30 RPBE, and 102 reported RPA calculations \cite{rpadata}. The fine-tuned model achieved a MAE of 0.034 eV on the training set and 0.036 eV on the RPA validation set (\textbf{Fig. ~\ref{Fig. 4}d} and \textbf{Fig. S4}). Hence, both scarce high-precision experimental and theoretical simulation data could be leveraged for the efficient prediction of more $E_{\rm{ad}}$ with chemical accuracy across various surfaces.  

%\section{Discussion}
%\textit{Discussion---}
In summary, established upon two key physical insights, the adsorption strength governed by orbital interactions and the functional-independent LDOS-adsorption relationship, the DOTA model permits the alignment of experimental sources and multi-fidelity DFT data. This allows achieving chemical accuracy in predicting $E_{\rm{ad}}$ across extensive metallic and intermetallic surfaces, primarily relying on GGA-level datasets augmented by minimum high-precision $E_{\rm{ad}}$ data. DOTA also exhibits exceptional transferability across chemical element environments. Thus, it tackles challenges of functional dependency and interpretability for deep learning models of $E_{\rm{ad}}$ prediction and limited availability of high-precision data in surface chemistry. It could provide high-precision $E_{\rm{ad}}$ data to accelerate the discovery of improved materials in heterogeneous catalysis, energy storage and conversion, and gas storage and separation. \par

\hspace*{\fill} 
\begin{acknowledgments}
%\section{Acknowledgments}
This work was supported by the National Key Research and Development Program of China (2023YFA1507601 and 2021YFA1500700). We thank the Centre for High Performance Computing at Shanghai Jiao Tong University for providing the computing resources of the Siyuan-1 cluster.
\end{acknowledgments}
%\hspace*{\fill} \\
%  Z.Zhang wrote the code, performed theoretical simulations, and wrote the paper. X.-M.C. designed the study, analyzed the data, and wrote the paper. Both authors discussed the results and improved the manuscript.\par
%\hspace*{\fill} \\
%All authors declare no competing interests.\par
\hspace*{\fill} \\
  %\textit{Data availability---}
The processed data used in this work are available via FigShare at \cite{DOTA_Data}. Code for DOTA is available as an open-source repository on GitHub at \cite{DOTA}
\nocite{VASP_2, PAW, ASE_2, CRC}
\bibliography{reference}
\bibliographystyle{apsrev4-2}

\end{document}

%\preprint{}

\title{Supplementary Materials for\\ \textbf{Orbital-interaction-aware Deep Learning Model for Efficient Surface Chemistry Simulations}}


\author{Zhihao Zhang}
\noaffiliation
\author{Xiao-Ming Cao}
\email[Contact author:]{xmcao@sjtu.edu.cn}
\noaffiliation
\date{\today}

%\keywords{}

\maketitle
This PDF file includes:\par
Materials and Methods\par
Figures S1 to S4\par
Tables S1 to S4

\section{Materials and Methods}
All the DFT calculations were performed with Vienna ab initio Simulation Package (VASP) \cite{VASP_2} with periodic boundary and the projector augmented wave (PAW) pseudopotentials \cite{PAW}. The Catalysis Kit (CatKit) software \cite{8_2} and the Atomic Simulation Environment (ASE) Python library \cite{ASE_2} were used to generate and classify the adsorption structures. The chemisorption structures were obtained in reference \cite{8_3} for the DOTA model pretraining, where the PBE functional was used for both $E_{\rm{ad}}$ and DOS profile calculations. For the fine-tuning stage, PBE, RPBE, and PBEsol functionals were used for $E_{\rm{ad}}$ and DOS profile calculations. The hybrid HSE06 functional was used only for the DOS profile of the gaseous adsorbate. All calculations were conducted with spin polarization, using a kinetic energy cutoff of 450~eV. The \textit{p}(2$\times$2) 3-layer slab was employed for each metallic and intermetallic surface. The corresponding Monkhorst-Pack k-point grid was 7$\times$7$\times$1, with a first-order Methfessel-Paxton smearing of 0.1~eV. The resolution of calculated lm- and site-DOSs with reference to the Fermi level, which were employed for DOTA training, was 0.01~eV. For dissociative adsorption, the $E_{\rm{ad}}$ was computed as $E_{ad}(X_2) = 2 \times E_{ad}(X) + D(X-X)$, where $E_{ad}(X)$ (X = H or O) was predicted by DOTA from atomic LDOS inputs, and $\text{D(H–H})$ or $\text{D(O=O)}$ denotes the standard gas-phase bond dissociation energy \cite{CRC}.\par
The DOTA model was implemented with the PyTorch Python library. The AdamW optimizer with an initial learning rate of 0.0002 and the L$\mathrm{_1}$ loss were used for training. An exponential cosine annealing scheduler was implemented to adjust the learning rate, with 150 epochs used, comprising 50 epochs for warming up. A batch size of 256 was employed. The contribution of each energy level to the $E_{\rm{ad}}$ was calculated using the interpretable integrated gradients algorithm implemented in the Captum library \cite{22_1} with default parameters.\par

\clearpage
\begin{figure*}
    \includegraphics{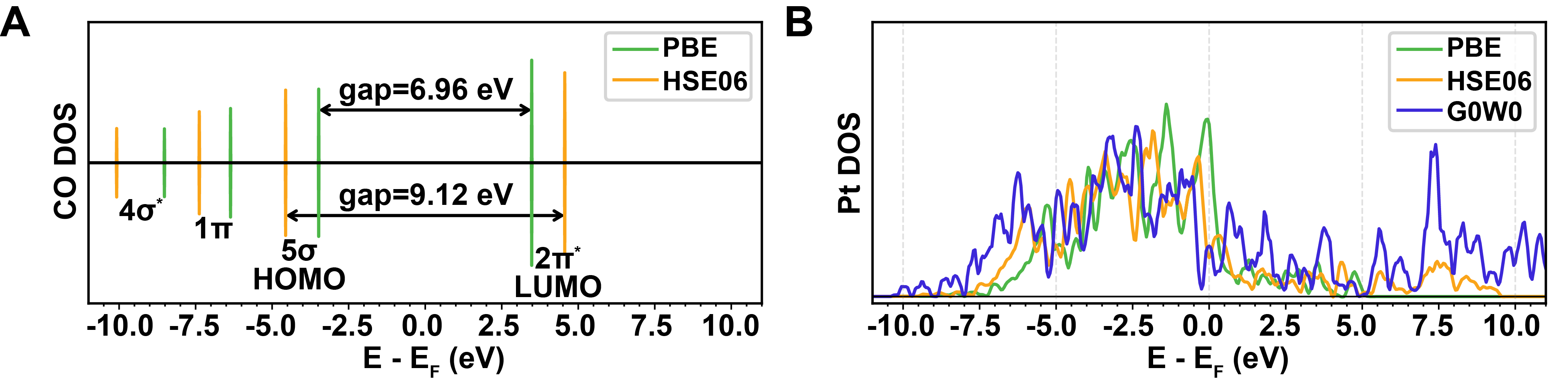} 
    \caption{\label{Fig S1}a) Density of states of CO obtained from different methods. b) Density of states of Pt slab obtained from different methods.}
\end{figure*}

\begin{figure*}
    \includegraphics{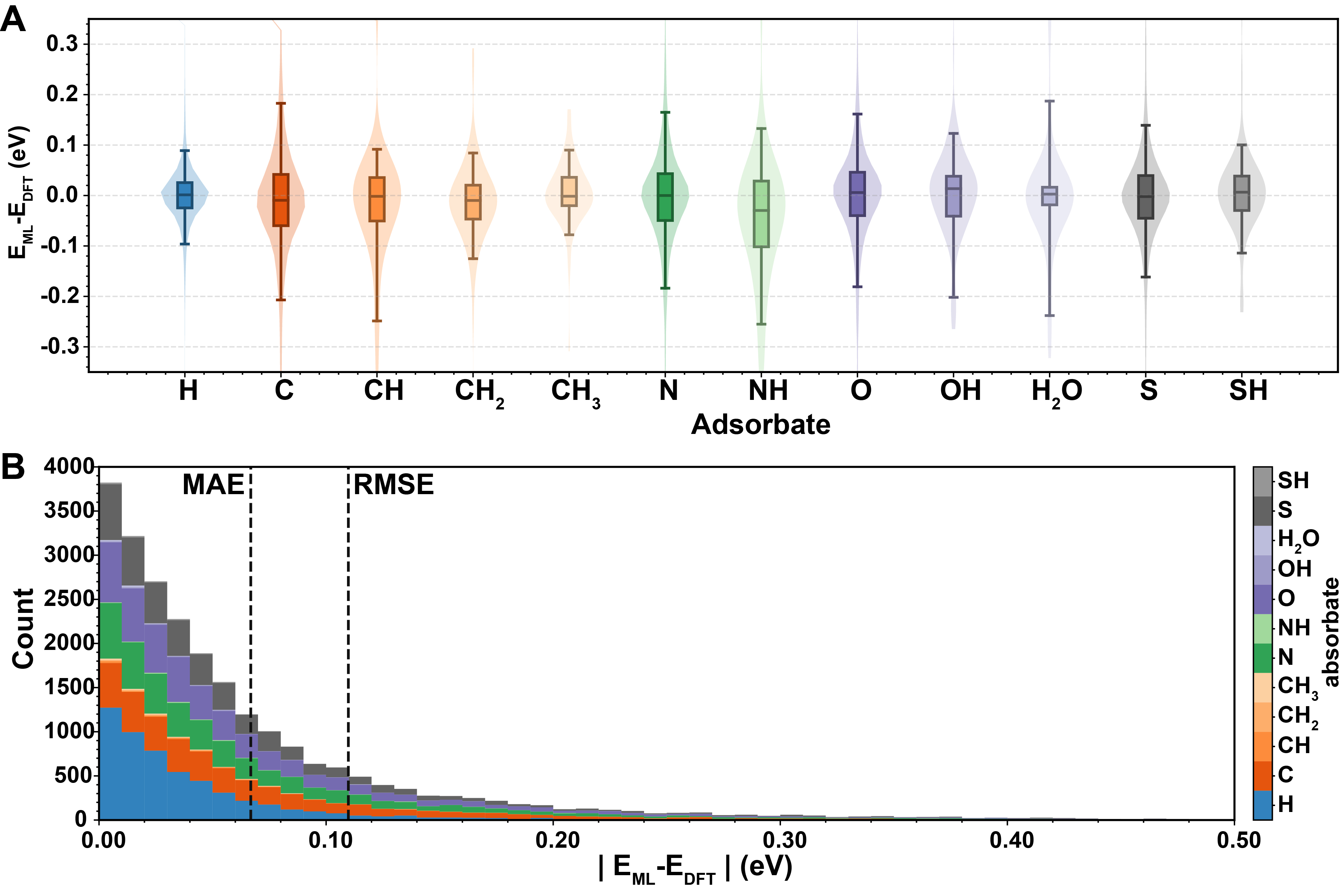}
    \caption{\label{Fig S2}a) Distribution of prediction errors for PBE adsorption energies across all the tested systems. Shown as violin plots (density distribution) overlaid with box plots (median, quartiles). The box spans the interquartile range (25th-75th percentiles), with the median indicated by the central line. Whiskers extend to the 5th and 95th percentiles. b) Histogram of prediction residuals.}
\end{figure*}
\clearpage

\begin{figure*}
    \includegraphics{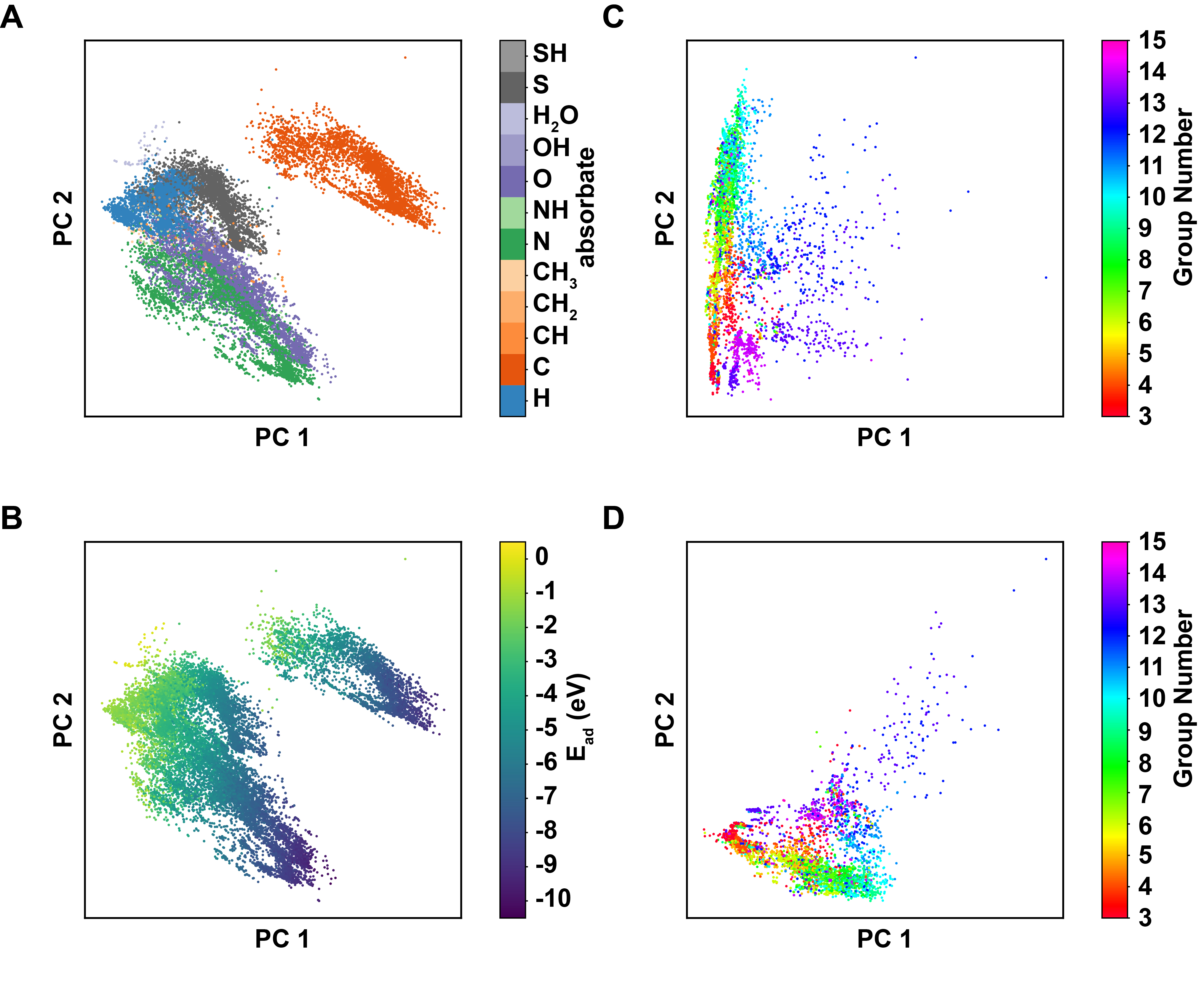}
    \caption{\label{Fig S3}a) Plot of the first component (PC1) versus the second component (PC2) of learned features from a principal component analysis for all pretraining systems, colored by adsorbate. b) Plot of PC1 versus PC2 for all pretraining systems, colored by adsorption energy. c) Plot of PC1 versus PC2 for H-adsorption systems, colored by host metal group number. d) Plot of PC1 versus PC2 for O-adsorption systems, colored by host metal group number. All features are extracted from the penultimate layer of the pretrained DOTA-PBE model. Structured clustering reveals that the model learns representations encoding adsorbates, energetic trends, and periodic dependence on metal groups without explicit supervision on these variables.}
\end{figure*}
\clearpage

\begin{figure*}
    \includegraphics{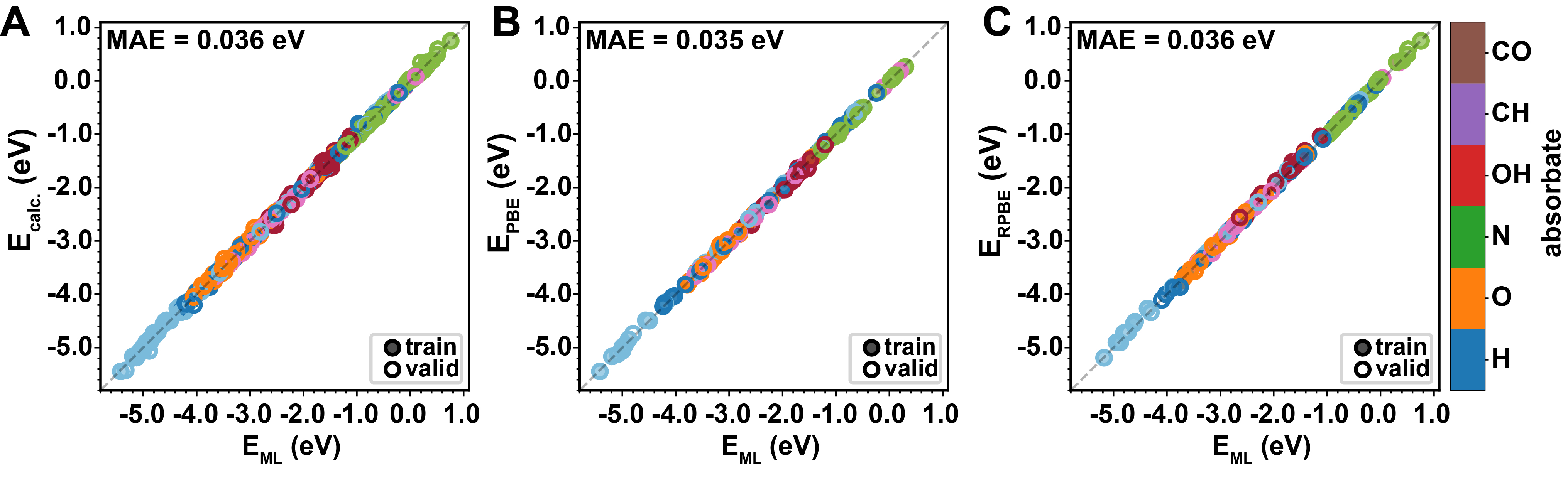}
    \caption{\label{Fig S4}a) DOTA-predicted adsorption energies at 100\% monolayer coverage versus reference data for multiple adsorbates across diverse metal surfaces of multiple methods (PBE, RPBE, RPA). b) DOTA-predicted adsorption energies at 100\% monolayer coverage versus PBE reference data for multiple adsorbates across diverse metal surfaces. b) DOTA-predicted adsorption energies at 100\% monolayer coverage versus RPBE reference data for multiple adsorbates across diverse metal surfaces.}
\end{figure*}

%%%%%%%%%%%%%%%% SUPPLEMENTARY TABLES %%%%%%%%%%%%%%%

\begin{table*}
    \caption{\label{table S1}Performance of DOTA for different adsorbates
    }
    \begin{ruledtabular}
    \begin{tabular*}{\textwidth}{
        @{\extracolsep{\fill}}
        l
        *{6}{c}
    }
    %\\
    %\hline
    \textbf{Adsorbate} & \textbf{Counts} & \textbf{Average}\newline \textbf{(eV)}
    & \textbf{SD}\newline \textbf{(eV)} & \textbf{MAE}\newline \textbf{(eV)} & \textbf{MAPE}\newline \textbf{($\boldsymbol{\rm \%}$)} & \textbf{RMSE}\newline \textbf{(eV)} \\
    \hline
    \textbf{H} & 5453 & -2.579 & 0.707 & 0.042 & 1.747 & 0.075 \\ 
    \textbf{C} & 4197 & -6.376 & 1.715 & 0.083 & 1.322 & 0.135 \\ 
    \textbf{CH} & 107 & -5.232 & 1.710 & 0.082 & 1.522 & 0.133 \\ 
    $\boldsymbol{\rm CH_2}$ & 82 & -3.808 & 1.340 & 0.060 & 1.654 & 0.089 \\
    $\boldsymbol{\rm CH_3}$ & 75 & -1.933 & 0.761 & 0.044 & 3.440 & 0.063 \\
    \textbf{N} & 4418 & -6.058 & 1.971 & 0.074 & 1.340 & 0.115 \\
    \textbf{NH} & 104 & -5.238 & 1.948 & 0.096 & 1.775 & 0.135 \\
    \textbf{O} & 4940 & -6.293 & 1.871 & 0.071 & 1.190 & 0.115 \\
    \textbf{OH} & 98 & -3.810 & 1.344 & 0.079 & 2.307 & 0.111 \\
    $\boldsymbol{\rm H_2O}$ & 60 & -0.417 & 0.368 & 0.061 & 49.632 & 0.114 \\
    \textbf{S} & 4227 & -5.195 & 1.212 & 0.066 & 1.307 & 0.106 \\
    \textbf{SH} & 100 & -2.427 & 1.097 & 0.053 & 2.664 & 0.078 \\
    \hline
    \textbf{Total} & \textbf{23861} & \textbf{-5.148} & \textbf{2.172} & \textbf{0.066} & \textbf{1.534} & \textbf{0.109} \\
    %\hline
    \end{tabular*}
    \end{ruledtabular}
\end{table*}

\begin{table*}
    \caption{\label{table S2}Data used for training DOTA
    }
    \begin{ruledtabular}
    \setlength{\tabcolsep}{1pt}
    \begin{tabular*}{\textwidth}{
        l
        c
        c
        c
        c
    }
    %\\
    %\hline
    \begin{tabular}{l}
         \textbf{Adsorbate}
    \end{tabular} &
    \begin{tabular}{c}
         \textbf{Counts}
    \end{tabular} &
    \begin{tabular}{c}
         \textbf{Functional}\\ \textbf{for DOS}
    \end{tabular} &
    \begin{tabular}{c}
         \textbf{Functional}\\ \textbf{for Energy}
    \end{tabular} &
    \begin{tabular}{c}
         \textbf{Surfaces}
    \end{tabular}\\
    \hline
    \begin{tabular}{l}
         \textbf{H}\\ \textbf{C}\\ \textbf{CH}\\ $\boldsymbol{\rm CH_2}$\\ $\boldsymbol{\rm CH_3}$\\ \textbf{N}\\ \textbf{NH}\\ \textbf{O}\\ \textbf{OH}\\ $\boldsymbol{\rm H_2O}$\\ \textbf{S}\\ \textbf{SH}
    \end{tabular} &
    \begin{tabular}{c}
         5453\\ 4197\\ 107\\ 82\\ 75\\ 4418\\ 104\\ 4940\\ 98\\ 60\\ 4227\\ 100
    \end{tabular} &
    \begin{tabular}{c}
         PBE
    \end{tabular} &
    \begin{tabular}{c}
         PBE
    \end{tabular} &
    \begin{tabular}{c}
         metallic and bimetallic alloy (111) surfaces \\
        with the combination of 37 metals \\
        (Sc, Ti, V, Cr, Mn, Fe, Co, Ni, Cu, Zn, \\
        Y, Zr, Nb, Mo, Tc, Ru, Rh, Pd, Ag, Cd, \\
        La, Hf, Ta, W, Re, Os, Ir, Pt, Au, Hg, \\
        Al, Ga, In, Sn, Tl, Pb, Bi)\\
    \end{tabular} \\
    \hline
    \begin{tabular}{l}
         \textbf{CO}\\ \textbf{CO}\\ \textbf{CO}
    \end{tabular} &
    \begin{tabular}{c}
         998\\ 998\\ 998
    \end{tabular} &
    \begin{tabular}{c}
         PBE\\ RPBE \\ PBESol
    \end{tabular} &
    \begin{tabular}{c}
         PBE\\ RPBE \\ PBESol
    \end{tabular} &
    \begin{tabular}{c}
         metallic and bimetallic alloy (111) surfaces \\
        with the combination of 11 metals \\
        (Co, Ni, Cu, Rh, Pd, Ag, Ir, Pt, Au, Sn, Pb)
    \end{tabular} \\
    \hline
    \begin{tabular}{l}
         \textbf{CO}
    \end{tabular} &
    \begin{tabular}{c}
         4
    \end{tabular} &
    \begin{tabular}{c}
         PBE (slab)\\ +\\ HSE06\\ (adsorbate)
    \end{tabular} &
    \begin{tabular}{c}
         Exp.
    \end{tabular} &
    \begin{tabular}{c}
         Pd(111)(fcc), \\ Cu(111)(top),\\ Ir(111) (fcc),\\ Ni(111)(fcc)
    \end{tabular}\\
    %\hline
    \end{tabular*}
    \end{ruledtabular}
\end{table*}

\begin{table*}
    \caption{\label{table S3}
		Adsorption energies of CO on Pt(111) predicted by models fine-tuned with the same set of high-fidelity experimental data points but different combinations of exchange-correlation functionals for LDOS construction.
    }
    \begin{ruledtabular}
    \setlength{\tabcolsep}{4pt}

    \begin{tabular*}{1.0\textwidth}{
         c
         c
         c
         c
         c
         c
        }
    %\\
    %\hline
    \begin{tabular}{c}
         \textbf{Functional for}\\ \textbf{adsorbate DOS}
    \end{tabular} &
    \begin{tabular}{c}
         \textbf{Functional for}\\ \textbf{surface DOS}
    \end{tabular} &
    \begin{tabular}{c}
         \textbf{$\boldsymbol{\rm E_{top}}$}\\ \textbf{(eV)}
    \end{tabular} &
    \begin{tabular}{c}
         \textbf{$\boldsymbol{\rm E_{fcc}}$}\\ \textbf{(eV)}
    \end{tabular} &
    \begin{tabular}{c}
         \textbf{preferred}\\ \textbf{site}
    \end{tabular} &
    \begin{tabular}{c}
         \textbf{$\boldsymbol{\rm E_{prefer}}$}\\ \textbf{(eV)}
    \end{tabular} \\
    \hline
        HSE06 & RPBE & -1.42 & -1.34 & top & -1.42 \\
        HSE06 & PBE & -1.37 & -1.25 & top & -1.37 \\
        HSE06 & PBESol & -1.42 & -1.34 & top & -1.42 \\
        RPBE & RPBE & -1.31 & -1.37 & fcc & -1.37 \\
        RPBE & PBE & -1.39 & -1.38 & top & -1.39 \\
        RPBE & PBESol & -1.40 & -1.41 & fcc & -1.41 \\
        PBE & RPBE & -1.37 & -1.46 & fcc & -1.46 \\
        PBE & PBE & -1.41 & -1.46 & fcc & -1.46 \\
        PBE & PBESol & -1.46 & -1.42 & top & -1.46 \\
        PBESol & RPBE & -1.45 & -1.44 & top & -1.45 \\
        PBESol & PBE & -1.46 & -1.47 & fcc & -1.47 \\
        PBESol & PBESol & -1.63 & -1.62 & top & -1.63 \\
    \hline
        exp. & exp. & & & top & -1.37$\boldsymbol{\rm \pm}$0.13\\
    %\hline
    \end{tabular*}
    \end{ruledtabular}
\end{table*}

\begin{table*}
    \caption{\label{table S4}
		Adsorption energies of CO on Rh(111) predicted by models fine-tuned with the same set of high-fidelity experimental data points but different combinations of exchange-correlation functionals for LDOS construction.
    }
    \begin{ruledtabular}
    \setlength{\tabcolsep}{4pt}

    \begin{tabular*}{1.0\textwidth}{
         c
         c
         c
         c
         c
         c
        }
    %\\
    %\hline
    \begin{tabular}{c}
         \textbf{Functional for}\\ \textbf{adsorbate DOS}
    \end{tabular} &
    \begin{tabular}{c}
         \textbf{Functional for}\\ \textbf{surface DOS}
    \end{tabular} &
    \begin{tabular}{c}
         \textbf{$\boldsymbol{\rm E_{top}}$}\\ \textbf{(eV)}
    \end{tabular} &
    \begin{tabular}{c}
         \textbf{$\boldsymbol{\rm E_{fcc}}$}\\ \textbf{(eV)}
    \end{tabular} &
    \begin{tabular}{c}
         \textbf{preferred}\\ \textbf{site}
    \end{tabular} &
    \begin{tabular}{c}
         \textbf{$\boldsymbol{\rm E_{prefer}}$}\\ \textbf{(eV)}
    \end{tabular} \\
    \hline
        HSE06 & RPBE & -1.52 & -1.43 & top & -1.52 \\
        HSE06 & PBE & -1.46 & -1.32 & top & -1.46 \\
        HSE06 & PBESol & -1.47 & -1.42 & top & -1.47 \\
        RPBE & RPBE & -1.54 & -1.41 & top & -1.54 \\
        RPBE & PBE & -1.52 & -1.45 & top & -1.52 \\
        RPBE & PBESol & -1.54 & -1.45 & top & -1.54 \\
        PBE & RPBE & -1.55 & -1.50 & top & -1.55 \\
        PBE & PBE & -1.57 & -1.50 & top & -1.57 \\
        PBE & PBESol & -1.60 & -1.47 & top & -1.60 \\
        PBESol & RPBE & -1.64 & -1.67 & fcc & -1.67 \\
        PBESol & PBE & -1.76 & -1.56 & top & -1.76 \\
        PBESol & PBESol & -1.75 & -1.64 & top & -1.75 \\
    \hline
        exp. & exp. & & & top & -1.45$\boldsymbol{\rm \pm}$0.14\\
    %\hline
    \end{tabular*}
    \end{ruledtabular}
\end{table*}
\clearpage

\bibliography{reference}
\bibliographystyle{apsrev4-2}

%%%%%%%%%%% CAPTIONS FOR OTHER SUPPLEMENTARY FILES %%%%%%%%%%